\documentclass{jpp}
\usepackage{graphicx}

\usepackage[utf8]{inputenc}
\usepackage[T1]{fontenc}
\usepackage{amsmath}
\usepackage{color}

\shorttitle{Current sheets in KAW turbulence}
\shortauthor{J Sharma, CA Kumar, KD Makwana, T N Parashar and S Satyasmita}

\title{Sub-ion scale current sheets in kinetic Alfv{\'e}n wave turbulence}

\author{Johan Sharma\aff{1}
  \corresp{\email{ph21resch11002@iith.ac.in}},
  Ch Akshath Kumar\aff{1}, Kirit D. Makwana\aff{1}, Tulasi N Parashar\aff{2}
 \and Sruti Satyasmita\aff{3}}

\affiliation{\aff{1}Department of Physics, Indian Institute of Technology Hyderabad, Sangareddy, Telangana 502284, India
\aff{2}Victoria University of Wellington, Kelburn, Wellington 6012, New Zealand
\aff{3}Charles University, Faculty of Mathematics and Physics, Prague, Czech Republic
}

\begin{document}

\maketitle

\begin{abstract}
3D kinetic particle-in-cell (PIC) simulations are performed using the kinetic Alfv{\'e}n wave (KAW) eigenvector relations from a two-fluid model as initial conditions, in order to study turbulent fluctuations and intermittent structures at sub-ion and electron scales. Simulations with different ion-to-electron mass ratios are set up to investigate the role of electron scales in the formation of intermittent structures. We analyze the current sheet structures that develop in these simulations. Two algorithms, namely Breadth-First Search (BFS) and Density-Based Spatial Clustering of Applications with Noise (DBSCAN), are employed to determine the thickness, length, and width of the current sheets, and both methods are found to yield consistent results. The average current sheet thickness scales inversely with the square root of the ion-to-electron mass ratio, with values close to the electron skin depth ($d_e$), indicating the presence of electron-scale current sheets in the simulations. The widths and lengths of the current sheets show a weaker scaling with the mass ratio. The scale-dependent kurtosis reveals enhanced intermittency at electron scales, consistent with magnetosheath observations. Distributions of scale-dependent properties of the current sheets also align with the electron skin depth of the different simulations {and they lie within ranges observed in kinetic scale solar wind turbulence}. This study reveals the nature of sub-ion-scale current sheets in KAW turbulence and their role in dissipation.
\end{abstract}

\section{Introduction}
The nature of turbulence at kinetic scales has been a long standing problem in solar wind, magnetosheath and other astrophysical environments. Two main processes have been suggested to account for the damping and dissipation of turbulence in the solar wind. One is collisionless damping mediated by wave–particle interactions \citep{howes2008inertial, schekochihin2009astrophysical}. Different waves have been proposed, like kinetic Alfv{\'e}n waves, whistler waves, ion-cyclotron waves \citep{leamon1998contribution, sahraoui2009evidence, gary2012forward}. Within the collisionless wave–particle interaction paradigm \citep{howes2008kinetic, howes2008model, schekochihin2009astrophysical}, it is argued that kinetic scale turbulent fluctuations chiefly undergo electron Landau damping when the ion plasma beta satisfies $\beta_i \lesssim 1$. \citet{tenbarge2013collisionless} suggested that a turbulent cascade of kinetic Alfvén waves, terminated by collisionless Landau damping, can adequately account for the observed magnetic energy spectrum in the dissipation range of solar wind turbulence. The other involves energy conversion within or near small-scale current sheets \citep{dmitruk2004test, markovskii2011short, matthaeus2011needs, karimabadi2013coherent}. One prominent possibility is magnetic reconnection, which can occur within these turbulent current sheets. This process can not only dissipate fluctuations but also help pass energy further down to sub-ion scales \citep{ boldyrev2017magnetohydrodynamic, mallet2017disruption, cerri2017reconnection, loureiro2017collisionless, sahraoui2020magnetohydrodynamic}.

Shear Alfvén waves are a familiar feature of the solar wind \citep{Alfven1942, BelcherDavis1971}, and their interactions are usually described using incompressible MHD turbulence theories \citep{Matthaeus1983, Goldreich1995toward, Galtier2000weak, cho2000anisotropy, bhattacharjee2001random, verma2004statistical, Boldyrev2006,  makwana2020properties, schekochihin2022mhd}. As the turbulent energy moves to smaller and more anisotropic scales \citep{ShebalinMatthaeus1983}, the perpendicular size of the fluctuations approaches the ion gyro-radius. At these scales, ions begin to drift away from the magnetic field lines while electrons remain tied to them, introducing finite-Larmor-radius effects and giving rise to kinetic Alfvén waves (KAWs) \citep{hasegawa1982alfven}. KAWs are widely considered important for heating and accelerating the solar wind, and they are often used to explain the power-law spectra observed below ion scales. Many recent studies support the idea of a KAW cascade in this kinetic range \citep{voitenko2004excitation, AlexandrovaSaur2009, sahraoui2009evidence, sahraoui2010three, shaikh2009whistler, howes2011gyrokinetic, boldyrev2012spectrum, chen2017nature, cerri2019kinetic, passot2019imbalanced, makwana2023linear}. Some studies suggest that KAWs are critically damped well before reaching electron scales \citep{podesta2010kinetic, verscharen2020dependence}. In such a situation, a cascade of whistler waves could become dominant close to electron scales \citep{shaikh2009spectral, GarySmith2009, gary2012forward, saito2012beta}. Comparisons between spacecraft measurements and theoretical predictions generally show that near ion scale solar-wind fluctuations behave more like KAWs than whistlers \citep{ salem2012identification, sahraoui2009evidence, sahraoui2010three, chen2013nature}.

\citet{karimabadi2013coherent} presented kinetic simulations that capture dynamics from macroscopic fluid scales down to electron-scale. They observed that the turbulent cascade gives rise to coherent structures, manifested as current sheets, which sharpen down to electron scales and produce intense, localized plasma heating. \citet{tenbarge2013current} found that the wave-driven turbulence self-generates current sheets, whose filling fraction is closely linked to the electron heating rate. \citet{wan2016intermittency} found most of the energy dissipation occurs in highly localized regions that occupy only a tiny fraction of the system’s volume. \citet{califano2020electron} also found that electron scale current sheets were formed in their 2.5D Hybrid-Vlasov-Maxwell simulations. \citet{azizabadi2021identification} performed $2$D hybrid kinetic simulations. They found the current sheets thin down to the scales below the ion inertial scales as much as the grid resolution allows. Their study reveals that the current sheets prefer to thin down below the ion inertial scale rather than to develop 2D tearing instability. By varying the mass ratio of ions and electrons, \citet{edyvean2024scale} showed that the current sheets get more localized for smaller $m_e/m_i$ values. The intensity of the current, as well as the maximum kurtosis of magnetic field increments, increase logarithmically with mass ratio. \citet{vega2023electron} also saw electron scale 2D current sheets in their simulation with fluid-kinetic spectral plasma solver, and \citet{sharma2019transition} used PIC simulations to study the transition from ion-dominated to electron-only reconnection. 

Electron-scale reconnecting current sheets were found by \citet{phan2018electron}, and kinetic scale current sheets were observed by \citet{satyasmita2024identifying} from the MMS observations, suggesting that electron scale reconnection plays a role in dissipating turbulent energy in astrophysical and space plasmas. \citet{lotekar2022kinetic} analysed the data from Parker Solar Probe at a distance of $0.2$ AU and \citet{vasko2022kinetic} from Wind spacecraft at $1$ AU from the sun and found kinetic scale current sheets of proton and subproton scales. They found that the current density, magnetic shear angles and current sheet amplitude in these sheets show scale dependent properties. They interpreted the scale dependence as strong evidence that solar wind turbulence produces kinetic scale current sheets. A coherent understanding of the sizes and intensities of current sheets in kinetic plasma turbulence is largely missing with the exception of a handful of studies.

Magnetohydrodynamic (MHD) and particle-in-cell (PIC) simulations of freely decaying plasma turbulence at moderate plasma $\beta$ were carried out by \citet{makwana2015energy} and they found an excellent level of agreement between the two approaches. In both models, thin current sheets emerged, similar to those observed in the solar wind. However, notable differences appeared at smaller scales. Fraction of total dissipation varied with the volume occupied by the current sheets, and PIC simulations produced broader, more diffuse current sheets compared to MHD. Direct measurements of sheet thickness confirmed that, in PIC, they form at scales comparable to the ion skin depth, whereas in MHD their thickness is effectively constrained by the numerical grid, indicating that PIC simulations capture the full cascade from fluid-like MHD scales down to kinetic scales. However, because of the use of electron-positron plasma, ion and electron skin depths could not be differentiated. This study was extended in \citet{makwana2017dissipation} for moderate and low $\beta$ plasma and it was found that the $J_{\parallel} \cdot E_{\parallel}$ term causes dissipation and heating of the particles. They showed that dissipation is localized in high current density regions with stronger $J_{\parallel}\cdot E_{\parallel}$ in low $\beta$ case. Numerical simulations and spacecraft observations show that the current sheets are the site of turbulence dissipation and heating \citep{wan2012intermittent, wan2015intermittent, chasapis2015thin, chasapis2018situ, navarro2016structure, greco2018partial, qudsi2020observations}. However, the processes responsible for this are not well understood.

In this paper, we perform fully kinetic 3D PIC simulations initialized with a superposition of several KAWs. Instead of random fluctuations or forcing, the exact KAW eigenvector relations from a 2-fluid model are used as the initial perturbation of density, velocity, electric, and magnetic fields in the kinetic PIC simulation. The electric and magnetic field polarization ratios from the simulation are compared with the analytical polarization ratios of KAW and whistler waves, and they match the KAW relation, indicating a KAW dominance at sub-ion scales. Simulations with varying mass ratios are performed while keeping the box size and injection wavenumbers fixed with respect to the ion skin depth. This makes the electron length scale move to smaller scales with increasing mass ratio and hence allows us to accurately study the effect of electron scales on the current sheet formation. Current sheets have been identified with two algorithms, namely BFS (Breadth First Search) and DBSCAN (Density-Based Spatial Clustering of Applications with Noise), and their length, width, and thickness are obtained using PCA (Principal Component Analysis) and convex hull methods. This allows a systematic analysis of the current sheet properties in KAW turbulence. Using the identified current sheets, their scale-dependent properties are measured and compared with recent measurements of solar wind turbulence. This comparison shows that kinetic scale current sheets are formed by turbulent fluctuations in the solar wind.

The structure of this paper is as follows: Section~2 describes the simulation setup. Here, we describe the evolution of turbulence with time and the magnetic energy spectrum in the perpendicular direction, also the perturbation ratios from the simulations are compared with the analytical results of KAWs and whistler waves. Section~3 describes the methods and algorithms used to identify the current sheet structures that form in the different simulations. Section~4 provides a two- and three-dimensional analysis of the current-sheet length, width, and thickness, and examines their dependence on the ion-electron mass ratio. Section~5 investigates the kurtosis and the scale-dependent properties of the current sheets, as well as their dissipation fractions . Finally, Section~6 summarizes our results and their implications.

\section{Simulation setup and identification of sub-ion scale fluctuations} \label{sec:simulation}
We perform our numerical simulations using the iPIC3D code - an implicit particle-in-cell (PIC) framework \citep{ markidis2010multi, lapenta2012particle}. The system is initialized with a superposition of kinetic Alfvén waves (KAWs) as described in \citet{sharma2024kinetic}, enabling the study of nonlinear interactions among wave modes in a collisionless plasma environment. The simulations employ normalized units: lengths are expressed in ion skin depth $d_i$, velocities in terms of the speed of light $c$, and time in units of the inverse ion plasma frequency $\omega_{p,i}^{-1}$, where $\omega_{p,i} = c/d_i$. Ion mass and charge are set to unity ($m_i = 1$, $e = 1$), and the number density is normalized to the background density $n_0$, yielding a normalized ion mass density of one. The magnetic and electric fields are both normalized to $c\sqrt{4\pi\rho_0}$, where $\rho_0$ denotes the total mass density. In these units, the Alfvén speed normalized to $c$ equals the normalized background magnetic field strength, $\hat{B}_0 = v_A/c = 0.01$. All linear eigenmode relations used for initialization are reformulated in this normalized framework. We employ a three-dimensional simulation domain with periodic boundary conditions in all directions. The background magnetic field $\mathbf{B}_0$ is initialized as uniform and aligned with the $z$-axis. Details of the domain size, grid resolution, and species thermal velocities are provided in Table ~\ref{tab:setups}.

The initial wave perturbations are constructed such that $k_\perp > k_\parallel$ (as the turbulence cascading from larger scales is expected to be anisotropic) for each mode and the total wavenumber satisfies $k d_i < 1$, corresponding to supra-ion scale fluctuations. {We have done simulations with different number of particles per cell per species, as shown in the Table \ref{tab:setups}}. The initial condition for any physical quantity $\psi$ - which includes the electric and magnetic fields, particle drift velocities, and densities - is given by:
\begin{align}
\psi(\mathbf{r}) = \psi_0 + \Re\left[\sum_{\mathbf{k}} \delta \psi e^{i\mathbf{k} \cdot \mathbf{r} + \phi_{\mathbf{k}}}\right],
\end{align}
where $\psi_0$ is the equilibrium value, $\delta \psi$ is the amplitude derived from the linear polarization relations, $\Re$ represents the real part of the expression, and $\phi_{\mathbf{k}}$ is a random phase assigned to each mode. A total of 16 wavevectors are used in the initial superposition. To maintain $k_\perp > k_\parallel$, the chosen modes satisfy $|n_z| < \sqrt{n_x^2 + n_y^2}$, where $n_{x,y,z}$ are the integer mode numbers associated with the wavevector components $k_{x,y,z} = 2\pi n_{x,y,z} / L_{x,y,z}$. The electric and magnetic field perturbations are imposed directly on the simulation grid. Corresponding density and velocity perturbations are incorporated into the initial particle distributions via shifted Maxwellians. For a species $s$ with temperature $T_s$, the thermal velocity $v_{\text{th},s}$ satisfies $k_B T_s = m_s v_{\text{th},s}^2 / 3$, and fluid velocities derived from the eigenmodes are included as a bulk drift.

We conduct three distinct simulations with varying ion-to-electron mass ratios, as summarized in Table~\ref{tab:setups}. The box size and the injection wavenumbers are fixed with respect to the ion skin depth, and the normalized ion mass remains unity. Thus, effectively we are reducing the electron skin depth scale as we increase the ion-to-electron mass ratio. The input control parameter is the amplitude of the electric field perturbation in the $x$-direction, $\delta E_x$, while all other field and particle perturbations are determined self-consistently from the two-fluid eigenmode structure. The simulation then evolves the fully nonlinear plasma dynamics from these initial conditions. There is negligible change in the magnetic spectrum and polarization ratios when we go from a cubical box with $480^3$ resolution to a box with $480\times 480\times 240$ resolution, reducing the resolution in the magnetic field parallel direction. Hence to reduce the computational resources, we have reduced the resolutions along $z$ to half of the resolution in the perpendicular directions. Additionally, in post-processing we used interpolation to adjust the number of boxes from $240$ to $480$ along the $z$-axis, ensuring the box is cubical for analysis, and found that this adjustment did not alter the results.

\begin{table}
  \begin{center}
\def~{\hphantom{0}}
  \begin{tabular}{lcccccccc}
 Name & $L_x,L_y,L_z$ & $m_r$ & Resolution & ppc per species & $v_{th,i}$ & $v_{th,e}$ & $\beta _{total}$ & $T_i/T_e$ \\[10pt]

 S0 & $20d_i$ & $25$ & $480\times 480\times 240$ & $5^3$ & $0.010288c$ & $0.050229c$ & $1.15$ & $1.049$\\ 
 
 S1 & $20d_i$ & $50$ & $480\times 480\times 240$ & $5^3$ & $0.010392c$ & $0.069282c$ & $1.13$ & $1.125$\\

 S2 & $20d_i$ & $100$ & $480\times 480\times 240$ & $5^3$ & $0.010436c$ & $0.102191c$  & $1.19$ & $1.043$\\

 S3 & $20d_i$ & $100$ & $560\times 560\times 280$ & $5^3$ & $0.010436c$ & $0.102191c$  & $1.19$ & $1.043$\\

 S4 & $20d_i$ & $50$ & $480\times 480\times 240$ & $6^3$ & $0.010392c$ & $0.069282c$ & $1.13$ & $1.125$\\

 S5 & $20d_i$ & $100$ & $480\times 480\times 240$ & $6^3$ & $0.010436c$ & $0.102191c$  & $1.19$ & $1.043$\\
  \end{tabular}
  \caption{Simulation parameters}
  \label{tab:setups}
  \end{center}
\end{table}

\begin{figure}
\centering
\includegraphics[width=1\linewidth]{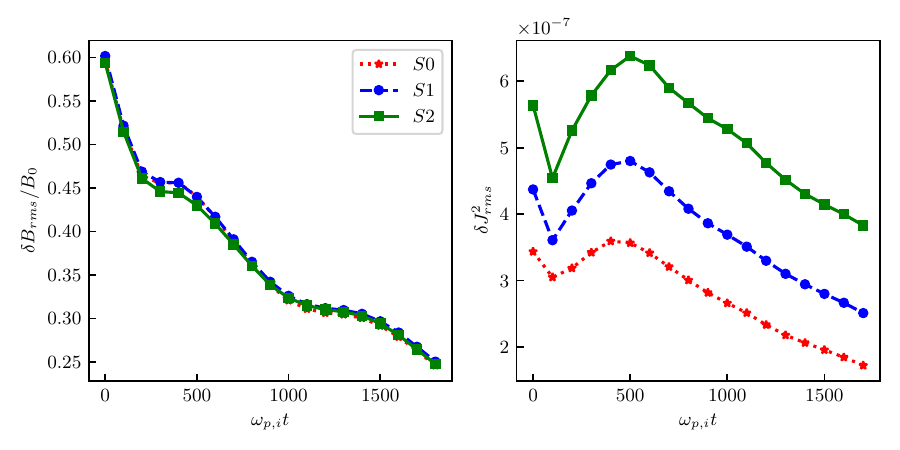}
\caption{$\delta B_{rms}/B_0$ (left) and $\delta J_{rms}^2$ (right) evolving with time for $S0$ (red), $S1$ and (blue) $S2$ (green) simulations.}
\label{fig:machno}
\end{figure}

We set up 3 simulations with different mass ratios ($S0$, $S1$, and $S2$ in table \ref{tab:setups}), and keeping approximately the same plasma $\beta$ ($\sim 1.1$) and also $T_i\approx T_e$ by adjusting the thermal velocities of ions and electrons. By studying the time evolution of $\delta B_{rms}/B_0$ for these 3 simulations, we look at the development of turbulence in the system and find the right time for analysis.
In Fig. \ref{fig:machno} we see the variation of $\delta B_{rms}/B_0$ with time for simulations with different mass ratios $S0$, $S1$, and $S2$. All 3 simulations show similar behavior with a maximum value of $\delta B_{rms}/B_0=0.59$ at $\omega _{p,i}t=0$ which then decreases with time, as is expected for a decaying simulation. Apparently, the mass ratio does not affect the time evolution of the average magnetic field fluctuation amplitude. The right panel of Fig. \ref{fig:machno} shows $\delta J_{rms}^2$ at different time steps. We see that the $\delta J_{rms}^2$ value peaks at $\omega _{p,i}t=400$ for $S0$ and, at $\omega _{p,i}t=500$ for $S1$ and $S2$ simulations, then decreases with time for all three simulations. The trend of the time evolution is similar across the three mass-ratios, however, the current density increases as the mass ratio increases. In MHD, the peak of $J^2$ would correspond to peak of dissipation and the phase after that as the freely decaying phase of turbulence. Hence, we choose the time period between $\omega _{p,i}t=600-1400$ for analysis of these current density structures. This ensures that the turbulence is fully developed, but the current sheets have not dissipated significantly.

\subsection{Magnetic spectrum}
\begin{figure}
\centering
\includegraphics[width=1\linewidth]{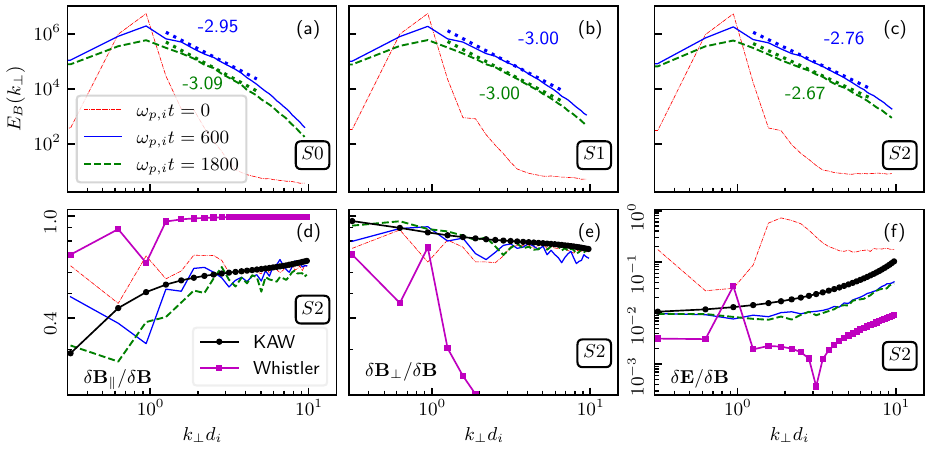}
\caption{ Magnetic spectrum in perpenducular wavevector $k_{\perp}d_i$ at $\omega _{p,i}t=0, 600$ and $1800$ for (a) $S0$, (b) $S1$ and (c) $S2$ simulation (top panel). Bottom panel shows the ratios (d) $\delta B_{\parallel}/\delta B$, (e) $\delta B_{\perp}/\delta B$ and (f) $\delta E/\delta B$ w.r.t $k_{\perp}d_i$ at $k_zd_i=0.314$, from $S2$ simulation, analytical results of KAW(black) and whistler wave(magenta).}
\label{fig:magspec}
\end{figure}

Magnetic wavenumber spectrum in the perpendicular direction ($k_{\perp}$) has been obtained by taking the discrete fourier transform of magnetic energy $|\hat{B}(k_{\perp},k_z)|^2 = |\hat{B}_x(k_{\perp},k_z)|^2 + |\hat{B}_y(k_{\perp},k_z)|^2 + |\hat{B}_z(k_{\perp},k_z)|^2$. Then, to obtain the magnetic spectrum in the perpendicular direction, we take a ring with a $k_{\perp}$ radius in the $k_x-k_y$ plane and obtain its average magnetic energy. Then other rings of different $k_{\perp}$ are formed, and a similar averaging operation is applied. Finally, the spectrum is summed over the $k_z$ direction and then multiplied with $k_{\perp}$ to obtain the magnetic energy along the perpendicular direction. Fig. \ref{fig:magspec} (a,b,c) shows the magnetic energy spectrum for the three simulations $S0$, $S1$ and $S2$ respectively at initial, intermediate and final time steps. At $\omega _{p,i}t=0$, only the lowest, initially excited, wavenumbers have energy. By $\omega _{p,i}t=600$, the turbulent cascade has spread energy across scales and we have fully developed sub-ion scale spectra with slopes of $-2.95$, $-3.00$ and $-2.76$ for $S0$, $S1$, and $S2$ simulations respectively. The slopes are obtained by choosing a $k_{min}d_i$ and $k_{max}d_i$ in each spectrum and fitting a line between these two points. The $k_{min}d_i$ is taken as $1$ for all the three simulations and $k_{max}d_i$ as $5$ for $S0$ and $7$ for $S1$, $S2$ simulations. At $\omega _{p,i}t=1800$ we get spectral slope values of $-3.09$, $-3.00$ and $-2.67$ respectively. As the mass ratio increases, the sub-ion scale spectral slope approaches $-2.67$, a number similar to previous simulations and observations in the solar wind and magnetosheath magnetic spectrum in the sub-ion range \citep{howes2008kinetic, matthaeus2008rapid, AlexandrovaSaur2009, sahraoui2010three, sahraoui2013scaling, karimabadi2013coherent, franci2015high, kobayashi2017three, matteini2017electric, parashar2018dependence, stawarz2019properties,  david2019spectrum}.

To investigate what kind of waves are present in these simulations, we obtain the different polarization ratios in the perpendicular wave vector from the simulation and compare them with the analytical results of KAWs and whistler waves taken from \citet{sharma2024kinetic, hollweg1999kinetic, swanson2003plasma}. To compute the ratios from the simulation, the magnetic and electric field data $\mathbf{B}(\mathbf{r})$ and $\mathbf{E}(\mathbf{r})$ are Fourier transformed to 
$\hat{\mathbf{B}}(\mathbf{k})$ and $\hat{\mathbf{E}}(\mathbf{k})$ at a chosen time. Then we take a small $k_{\perp}$ radius in $k_x - k_y$ plane at $k_zd_i=0.314$ with small $ dk_{\perp}$ to form a ring, in order to compare with the analytical relations that are also calculated at this $k_z$. The average of both numerator and denominator in this ring is obtained separately, and then the ratio is taken. Fig. \ref{fig:magspec}(d) shows the ratio $\delta B_{\parallel}/\delta B$ from the simulation $S2$ at initial, intermediate, and final time steps as well as from the analytical results of the KAW and the whistler waves (WW). We see that the simulation results do not match the analytical results of the whistler waves but are very close to the analytical KAW results. Similar results are found for the ratio $\delta B_{\perp}/\delta B$ in the Fig.~\ref{fig:magspec}(e), i.e., simulation results are close to analytical KAWs and far from WW analytical results. Fig.~\ref{fig:magspec}(f) shows that at $\omega _{p,i}t=0$ the simulation results of $\delta E/\delta B$ are far from both analytical whistler wave and analytic KAWs, but as the simulation evolves with time $\omega _{p,i}t=600, 1800$, the polarization ratios evolve to match closely with the analytical KAW expectation. These polarization ratios suggest that the KAWs are excited as the energy cascades below to sub-ion scales. Care, however, must be taken to interpret these results as we have not been able to simulate smaller electron masses to see the potentially critical damping of KAWs before reaching electron scales \citep{podesta2009dependence, verscharen2020dependence}.

\section{Current sheet detection}
We now describe the identification of coherent current sheets and their properties. First, the data is smoothed using a low pass Fourier space filter. This smoothing technique is applied to current density, velocity, charge density, electric, and magnetic field data. We also utilize the Savitsky-Golay filtering, and we describe later that the results do not show any qualitative change with different filtering choices. For the identification of current sheets, several algorithms have been used in the literature, which are similar in some ways but different in their exact methods. These tools have been applied in studies ranging from 3D MHD turbulence to 2D and 3D turbulence at kinetic scales, as well as in spacecraft measurements of the solar wind \citep{uritsky2010structures, zhdankin2013statistical, zhdankin2016scalings, servidio2015kinetic, sisti2021detecting, hu2020identifying, makwana2015energy, Khabarova_2021}. Here we broadly divide the entire methodology into two parts, (i) identifying the current sheets and (ii) calculating the dimensions of the current sheets. Identification of the current sheets is achieved using two algorithms in this study: Breadth-First Search (BFS) and Density-Based Spatial Clustering of Applications with Noise (DBSCAN). Principal Component Analysis (PCA) and convex hull is incorporated to calculate the dimensions of the current sheets identified. These will be briefly discussed in this section.  

\subsection{Identifying the Current Sheets}

\subsubsection{Breadth First Search}

We first take a $(2n+1) \times (2n+1)$ square (for 2D slices taken perpendicular to the mean magnetic field), or $(2n+1) \times (2n+1) \times (2n+1)$ cube for the 3D data, and traverse through the entire space. We label the point with maximum current density, $|J_z|$, in the square or cubical window as the local maxima. Now that we have identified the local maximum of each current sheet, we find the current sheet by the BFS Algorithm as the current sheet is a localized region with high current densities. From the local maximum, we initiate graph-based traversal via its neighboring points, four in the 2-Dimensional case and 6 in the 3-Dimensional case, two neighbors corresponding to each dimension. The traversal is continued till the current density in the neighboring location is above the threshold value $J_{tr}$ which is given by $J_{tr} = f \cdot J_{max}$, where $J_{max}$ is the current density value at the local maximum and $f$ is a parameter. In this study it has been set to $0.45$ for $2$D and $0.6$ for the $3$D case. If $J_{tr}$ i.e. $f$ is taken too large, then it misses many current density structures, and if it's too small, then it mistakenly identifies noise as a current sheet. In $3$D if we take $f=0.5$ then some of the the current sheets merge into one another. The search is continued till the current density drops below the threshold value. All the points that satisfy the above threshold condition are grouped as a current sheet. The parameters used to identify current sheets are carefully selected through visual inspection of the identified structures, ensuring that noise is excluded from the analysis while retaining all significant high current density regions. If the value of $n$ is too large, then it misses some current sheet structures, and if it is too small, then it identifies two neighbouring current sheets as a single current sheet. So we choose an optimal value of $n=15$ for $2$D and $40$ for $3$D. The same values of the parameters are used for analysis of all the simulations.

\subsubsection{Density-Based Spatial Clustering of Applications with Noise (DBSCAN)}
This is an unsupervised machine learning algorithm usually used to identify clusters in noisy data. DBSCAN has two main parameters, $\epsilon$, which is the radius to search for the neighbours, and minimum-samples $(M_n)$, meaning the minimum number of points needed to form a dense region. Firstly, we set a lower limit on the current density on the data we have, which is given by $J_{min}=1.2 J_{rms}$ for the $2$D and $2.5 J_{rms}$ for $3$D. All the points that lie below this threshold are filtered out from the data. The points that satisfy the threshold criteria form the points of our interest and DBSCAN is applied to these points. {To determine appropriate DBSCAN parameters, a $k$-distance graph was constructed by computing, for each data point, the distance to its $k^{th}$ nearest neighbor and sorting these distances in ascending order. The resulting curve was inspected to identify the elbow point, which was selected as the optimal range of $\epsilon$ in the filtered dataset \citep{ester1996dbscan}}. This method provided a data-driven starting range for the $\epsilon$ parameter. If the $\epsilon$ value is too small, then it misses some of the high current density regions, and if it's too large, then it includes noise as a cluster. Similarly, if $M_n$ is large, then it skips some current sheets, and if it's small, it includes noise. So we chose an optimal value for these parameters, with $\epsilon$ as 3.5 for 2D and 3.8 for 3D, and $M_n$ as $20$, $80$ in $2$D and $3$D analysis, respectively. 
To identify the clusters, the algorithm starts with an unvisited point say $p$ and searches all the points in its $\epsilon$ radius. If the number of points in this $\epsilon$ radius is greater than $M_n$, then the point is identified as a core point. All the points in the $\epsilon$ radius of the core point are added to the cluster, and if there is a second core point in its $\epsilon$ radius, then the second core point including the points in its $\epsilon$ radius are also added to the cluster, and this process continues to form large clusters. All points with fewer points than $M_n$ within their $\epsilon$ radius and not part of a cluster are labeled as noise. 
With the chosen parameters, DBSCAN was applied to the data to identify distinct clusters, with each cluster representing a unique current sheet. Both BFS and DBSCAN algorithms generate similar results, which will be discussed in the subsequent sections.

\subsection{Examples of identified current sheets}
\subsubsection{Examples in 2D}

\begin{figure}
\centering
\includegraphics[width=1\linewidth]{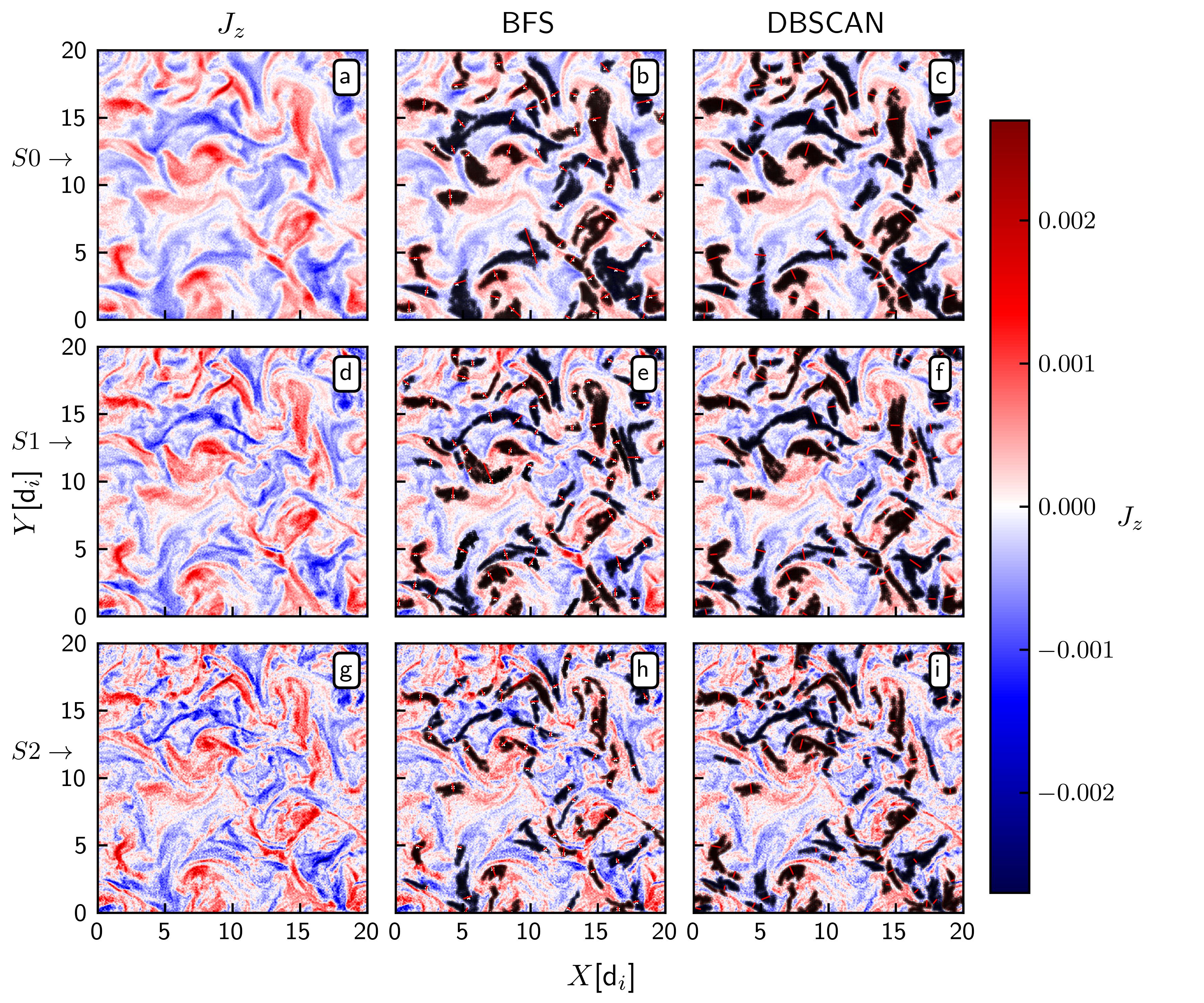}
\caption{ Current density $J_z$ in $x-y$ plane at $\omega _{p,i}t=600$ for S0 (a,b,c), S1(d,e,f) and S2(g,h,i) simulations. The current sheets identified with both BFS (middle panel) and DBSCAN (right panel) are shown in black clusters. The white cross lines shows the maxima of each cluster and solid red lines indicate the thickness of the identified clusters.}
\label{fig:2d_current_density}
\end{figure}

In this section we discuss the clusters identified in 2D slices taken perpendicular to the mean magnetic field. Fig. \ref{fig:2d_current_density} left panel shows the current density $J_z$ in $x-y$ plane with $z=0.042$d$_i$ at $\omega _{p,i}t=600$, middle panel showing the current sheets identified with BFS algorithm and right panel with DBSCAN. We see that both cluster identification methods, i.e., BFS and DBSCAN, identify almost the same clusters. Fig.  \ref{fig:2d_current_density} top row represents current sheets from $S0$ simulation, middle row from $S1$ and bottom row from $S2$ simulation. Almost similar current sheets are identified by the two algorithms except in some cases where they complement each other. For ex., the small current sheet on top left corner in Fig. \ref{fig:2d_current_density} (e,f) is identified by BFS Fig. \ref{fig:2d_current_density}(e) but not by DBSCAN Fig. \ref{fig:2d_current_density}(f). Another case where the current sheet in the bottom left corner of Fig. \ref{fig:2d_current_density} (e,f) is identified by DBSCAN Fig. \ref{fig:2d_current_density}(f) but not by BFS Fig. \ref{fig:2d_current_density}(e).

\subsubsection{Examples in 3D}

\begin{figure}
\centering
\includegraphics[width=1\linewidth]{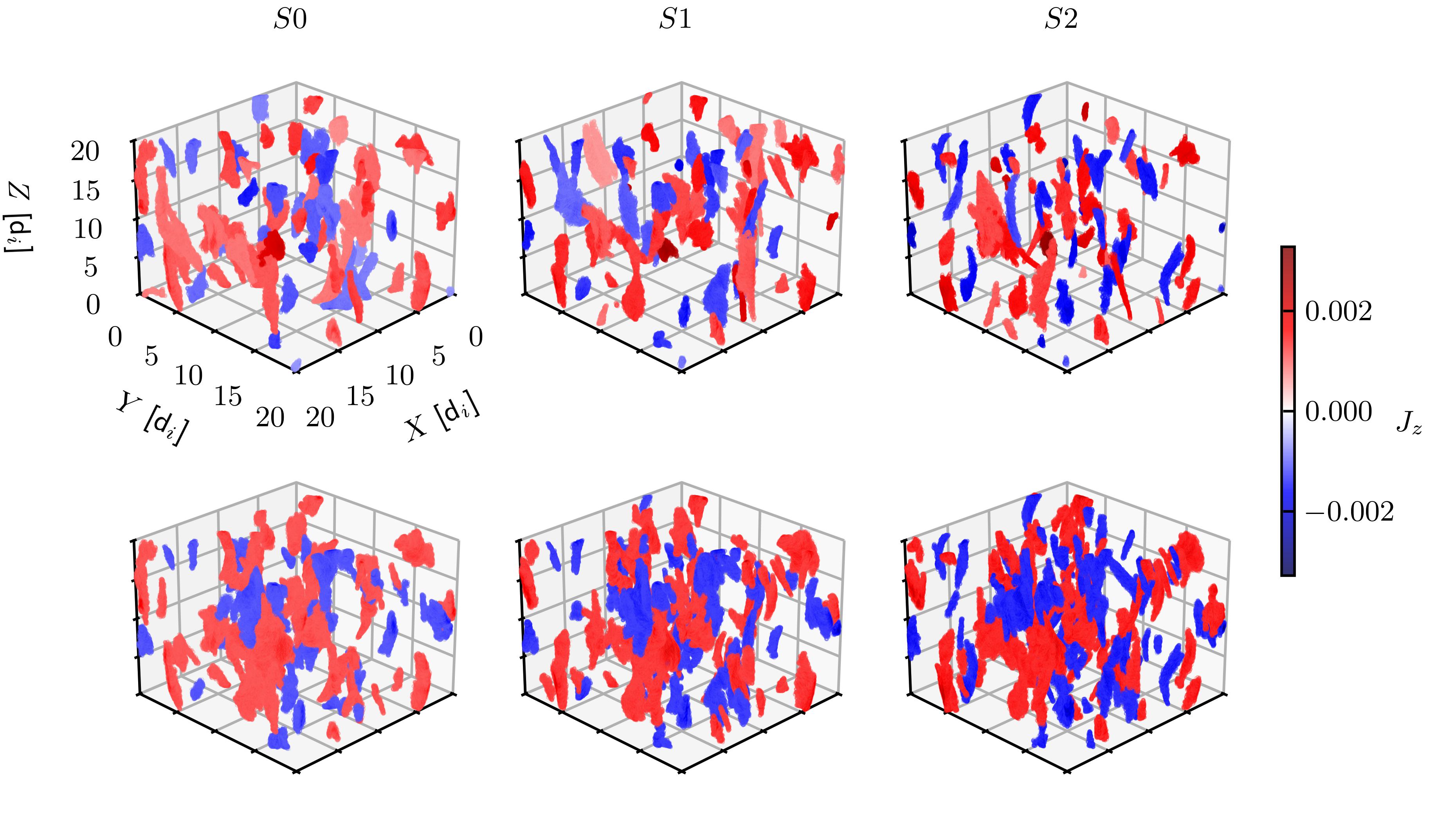}
\caption{ Current density $J_z$ structures in 3D at $\omega _{p,i}t=700$ for S0 ($m_r=25$), S1 ($m_r=50$) and S2 ($m_r=100$) simulations identified with BFS algorithm (top) and DBSCAN (bottom).}
\label{fig:3D_current_bfs}
\end{figure}

Here we present the current density structures identified in 3D analysis. Fig. \ref{fig:3D_current_bfs} shows the current sheets identified with BFS (top) and DBSCAN (bottom). This screenshot is from $S0$, $S1$, and $S2$ simulations from left to right at $\omega _{p,i}t=700$. Similar to 2D analysis, we see similar current sheets getting identified by the two methods, with some complentarity. Half-thickness, widths, and the lengths of the clusters are identified for the three simulations $S0$, $S1$, and $S2$.
Visual inspection of the simulation data provides insight into the geometry of the current sheets. 

\subsection{Dimensions of the current sheets}
\begin{figure}
    \centering
    \includegraphics[width=1\linewidth]{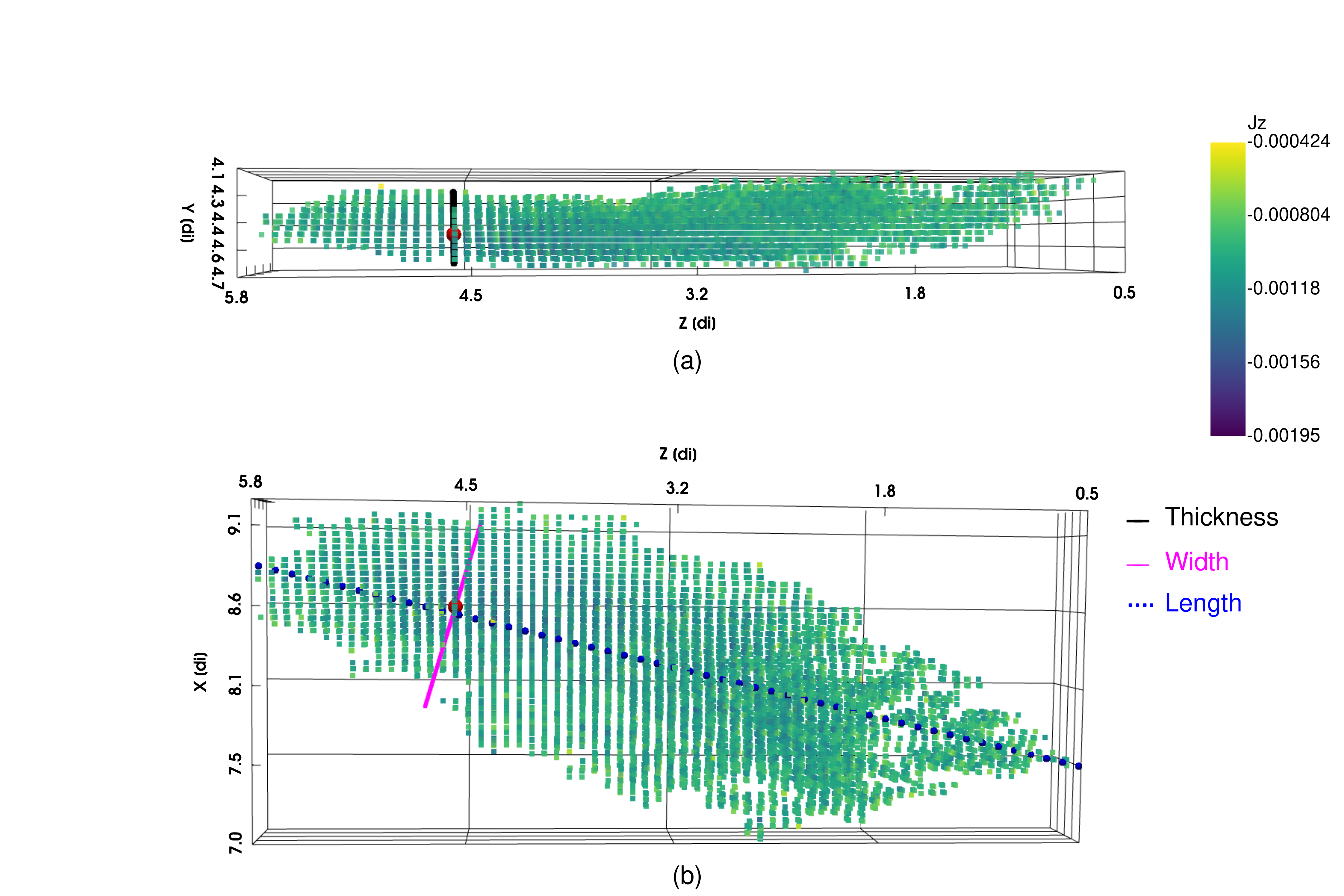}
    \caption{ Visualization of an isolated current sheet identified using BFS with the maximum current density point marked by a red sphere; (a) shows the thickness with a black line and (b) shows the width with a magenta line and the length with a blue dotted line.}
    \label{fig:all_three}
\end{figure}

Once we identify the current sheets, we calculate their dimensions, that is, the length, width, and thickness ($2 \times$half-thickness). To calculate the length of the current sheet in $3$D and the width in $2$D analysis, we identify the farthest points that are part of the particular current sheet. But if we check the distance of each pair possible, due to the enormous data, it takes a lot of computational power as the complexity of this algorithm is $\mathcal{O}(n^2)$ where $n$ is the number of points in a current sheet. To efficiently determine the maximum distance between any two points within a cluster, we leverage the property that these points must lie on the cluster's convex hull - {the smallest convex polygon that contains all points in the dataset.} This approach involves first constructing the convex hull of the point set, {which was done using \texttt{scipy.spatial.ConvexHull} \citep{virtanen2020scipy}}. The algorithm then finds the maximum distance by evaluating all pairs of vertices on the hull, which is computationally more efficient. The efficiency of this approach stems from the fact that the number of hull vertices, \(h\), is typically orders of magnitude smaller than the total number of cluster points, \(n\). This significant reduction in the number of candidate points for distance calculation reduces the computational complexity to \(\mathcal{O}(n\cdot \log (n)+h^{2})\), offering a substantial performance improvement over brute-force methods.
For the 3D current sheets, it is expected that the lengths of the current sheets identified with BFS to be slightly less than that of DBSCAN as the points which are being considered as a current sheet are restricted only till the 60{$\%$} of the maximum current density of the current sheets while in DBSCAN, it goes down till 50{$\%$}. {The identified current sheets are filtered such that each point within a current sheet maintains a current density that is at least 50{$\%$} of the maximum current density of that particular current sheet. Points that do not meet this criterion are excluded to ensure consistency in the dimensional analysis of current sheets}. 

To calculate the width and thickness of the current sheets in $3$D and thickness in $2$D, we apply Principal Component Analysis (PCA) on the current sheet. This gives us the directions along the length, width, and thickness of the current sheet, respectively. We initialize the local maxima as the starting point and traverse along the directions of width and thickness until we reach the half of the peak value of the current sheet. This is extended to the opposite direction as well to capture the full width and thickness of the current sheet. Fig.~\ref{fig:all_three} shows a typical current sheet identified. The sheet exhibits a complex 3D structure. Fig. \ref{fig:all_three}(a) shows the identified thickness, and Fig. \ref{fig:all_three}(b) shows the identified width and length of the current sheet.

\section{Current sheet statistics}
\subsection{Statistics in 2D}
\subsubsection{Halfthickness}


\begin{figure}
\centering
\includegraphics[width=1\linewidth]{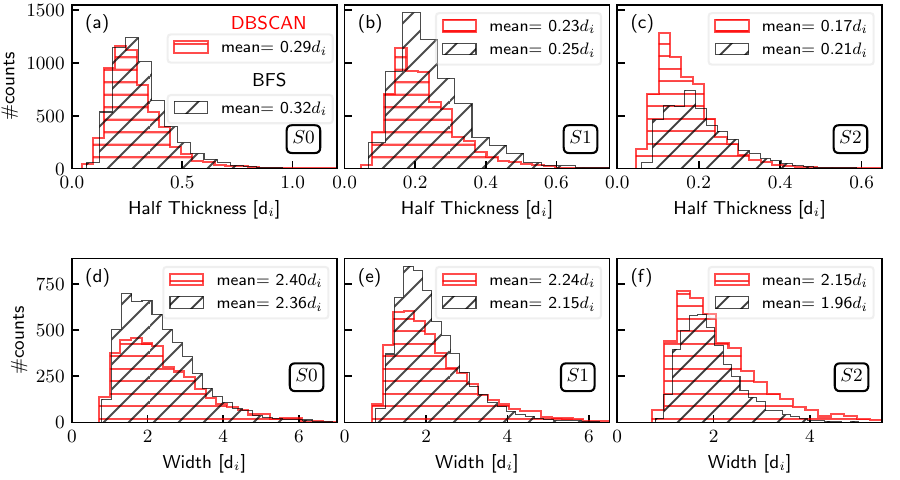}
\caption{ Distribution of half-thickness(top panel) and widths(bottom panel) of current sheets $J_z$ in $x-y$ plane, with 11 slices along z-axis at intervals $\Delta z=2d_i$ interval, over $\omega _{p,i}t=600-1400$ for $S0$ (a,d), $S1$ (b,e) and $S2$ (c,f) from left to right respectively. The red border histogram is obtained from DBSCAN and black border for BFS.}
\label{fig:2d_hist_width}
\end{figure}

The top row (a,b,c) in Fig. \ref{fig:2d_hist_width} shows the histograms of half-thickness of identified current sheets during the time period $\omega _{p,i}t=600-1400$ as identified by the two algorithms. Every time step consists of 11 slices of current density $J_z$ along z-axis at each $\Delta z=2d_i$ interval. The average values of the half-thickness from BFS for these 3 simulations are $0.32 d_i$, $0.25 d_i$, $0.21 d_i$ for mass ratios $25$, $50$ and $100$ respectively. The DBSCAN analysis gives very similar average thicknesses of $0.29 d_i$, $0.23 d_i$, $0.17 d_i$ for mass ratios $25$, $50$, and $100$, respectively. We see that similar to the previous simulations varying the mass ratio \citep{edyvean2024scale}, the half-thickness of the current sheets decreases with increasing mass ratio (or decreasing electron mass).

\subsubsection{Widths}

The bottom panels (d,e,f) of Fig. \ref{fig:2d_hist_width} show the histograms of the widths of current sheets identified between times $\omega _{p,i}t=600-1400$ by the two methods for $S0$, $S1$, and $S2$. Since convex hull is used for this calculation, all points below $0.5J_{max}$ are removed from the cluster and in DBSCAN, clusters with $J_{max}<2.4J_{rms}$ are ignored. Their averages are given by $2.36 d_i$, $2.15 d_i$ and $1.96 d_i$ from BFS and $2.40 d_i$, $2.24 d_i$ and $2.15 d_i$ from DBSCAN for $S0$, $S1$ and $S2$ simulation respectively. We get very similar results for both cases. We see that the widths decrease marginally with increasing mass ratio from $25$, $50$, to $100$, but the percentage decrease is not as substantial as for the thickness.

\subsection{Statistics in 3D}
\subsubsection{Halfthickness}


Fig. \ref{fig:3d_hist_lengths} top panel shows the histograms of half-thickness for 3D clusters obtained with BFS and DBSCAN for the three simulations for the time window $\omega _{p,i}t=600-1400$. The average value of half-thickness for both BFS and DBSCAN are $0.28 d_i$ ($1.4d_e$), $0.20 d_i$ ($1.4d_e$), $0.14 d_i$ ($1.4d_e$) for simulations $S0$, $S1$ and $S2$ respectively. The half-thickness decreases by a factor $\sqrt{2}$ for each increase of a factor of two in the mass ratio,  which is more pronounced than the 2D case. This means that the average thickness of the current sheets scales as the inverse of the square root of the mass ratio. Since increasing the mass ratio pushes the electron scale to smaller values, this indicates that the electron scale current sheets are being formed in these simulations.

\subsubsection{Widths and Lengths}

\begin{figure}
\centering
\includegraphics[width=1\linewidth]{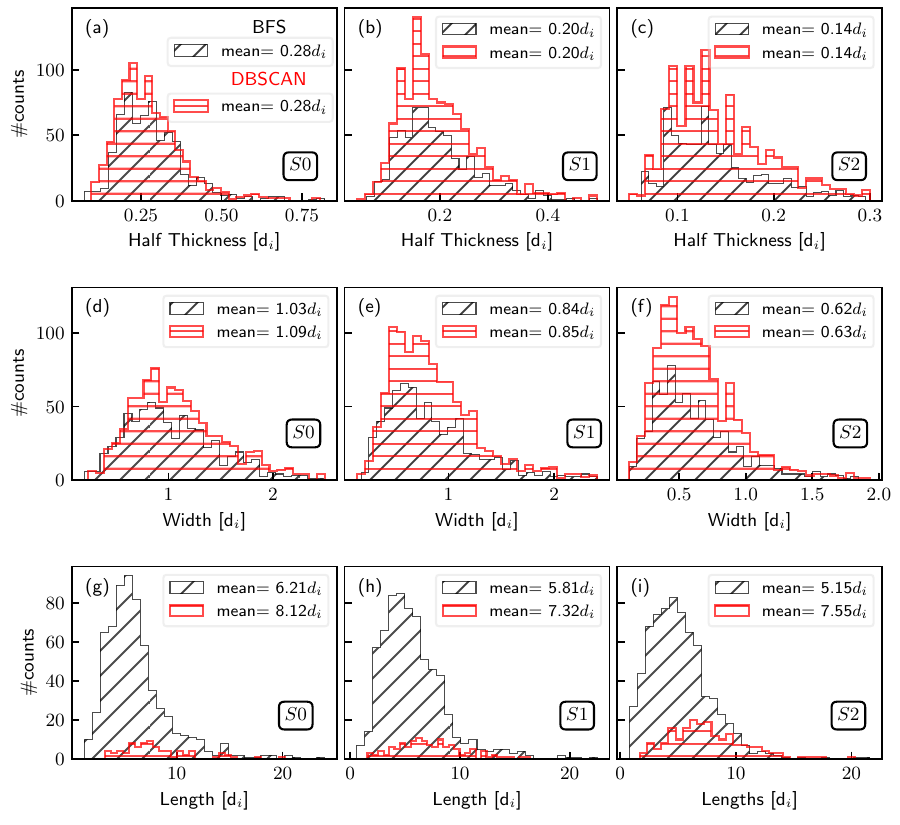}
\caption{ Half-thickness, width and length for current sheets $J_z$ in 3D measured over $\omega _{p,i}t=600-1400$ for $S0$ (a,d,g), $S1$ (b,e,h) and $S2$ (c,f,i) from left to right respectively. The red border histogram is obtained from DBSCAN, and the black border for BFS. The top panel shows half-thickness, the middle panel shows width, and the bottom panel shows length.}  \label{fig:3d_hist_lengths}
\end{figure}

\begin{figure}
\centering
\includegraphics[width=1\linewidth]{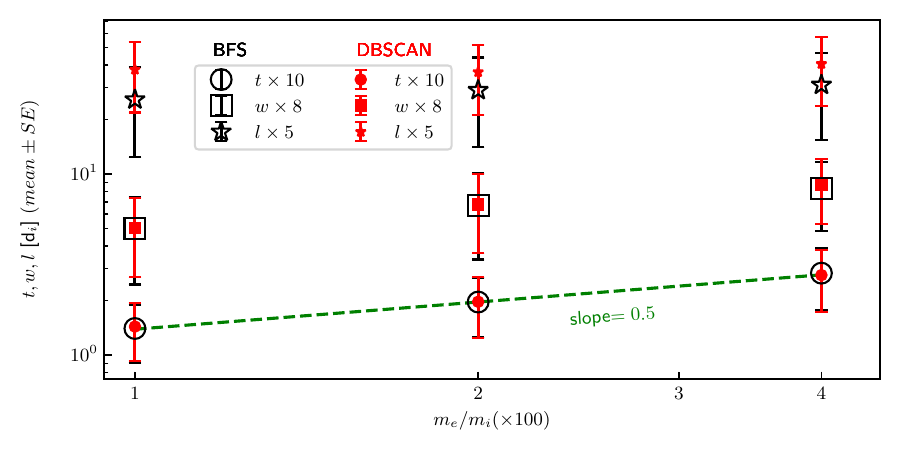}
\caption{ The mean of halfthickness($t$), width($w$) and length($l$) from $3$D statistics with error bars for $S0$, $S1$ and $S2$ simulation from BFS and DBSCAN, versus the mass ratio ($\times 100$ for visual clarity). The length, width, and half-thickness are multiplied by factors of $ 5$, $ 8$, and $10$, respectively, for a better visual representation.}
\label{fig:mean_sqrt}
\end{figure}

We also obtain the widths and lengths for the three simulations for both BFS and DBSCAN clusters. Similar to above cases, first a histogram of the distribution of the widths and lengths for a time $\omega _{p,i}t=600-1400$ is obtained and then the average value of that histogram is obtained. For BFS clusters we get average widths of $1.03 d_i$, $0.84 d_i$, $0.62 d_i$ for simulations $S0$, $S1$ and $S2$ respectively. Similarly for DBSCAN clusters we get widths $1.09 d_i$, $0.85 d_i$ and $0.63 d_i$ for the different simulations (see Fig. \ref{fig:3d_hist_lengths}(d,e,f)). The widths in 3D analysis are shorter that those obtained in 2D slice analysis. This is because in 3D we use the PCA direction to determine where the width falls below 50\% of the peak, but this may happen before the current sheet actually ends due to curvature. The convex hull method in 2D analysis does a better job of estimating the widths.

The average lengths are $6.21 d_i$, $5.81 d_i$, $5.15 d_i$ and $8.12 d_i$, $7.32 d_i$, $7.55 d_i$ for BFS and DBSCAN clusters for simulations $S0$, $S1$, and $S2$ respectively, as shown by the histograms in the bottom panels of Fig \ref{fig:3d_hist_lengths}. Convex hull method is used for length calculation because PCA will give inaccurate results due to the curvature in longer sheets. We see that the average widths of current sheets decrease with increasing mass ratio, but their scaling with the mass-ratio is weaker than that of the thicknesses. The lengths in BFS show slight decrease with mass-ratio, but there is no clear trend in the DBSCAN method, indicating that their scaling with mass-ratio is even weaker. The convex hull is used for length calculations, but the threshold for identifying cluster points in BFS is $0.6J_{max}$ while DBSCAN has a global threshold of $2.5J_{rms}$. In order to define the length consistently in DBSCAN, we remove all the points which are below $0.5J_{max}$ and also ignore clusters with $J_{max}<5J_{rms}$. Due to this the number of clusters for length calculations in DBSCAN are reduced compared to BFS. Also since the BFS clusters only extend up to $0.6J_{max}$, their lengths are shorter than DBSCAN lengths. Nevertheless, the scaling of lengths with mass-ratio appears to be weak in both analyzes. The mean of halfthickness($t$), width($w$) and length($l$) with error bars for $S0$, $S1$ and $S2$ simulation from BFS and DBSCAN is shown in Fig. \ref{fig:mean_sqrt} for visual comparison. We can see the clear $\sqrt{m_e/m_i}$ scaling of current sheet thicknesses, while the widths and lengths have weaker scaling but also larger error bars.

\subsection{Stability Under Numerical Variations}
To verify the robustness of our results, we analysed the current sheets with different scales for our low-pass FFT filter. In this paper, the FFT low-pass filtering is used with a cutoff at $k/k_{max}=0.5$. The FFT filtering with cut-off $k/k_{max}=0.625$ gives half-thicknesses of $0.36 d_i$, $0.25 d_i$, and $0.18 d_i$ for $S0$, $S1$, and $S2$ simulation respectively with BFS. By comparing these results with a cutoff $k/k_{max}=0.5$, we see the higher cutoff gives slightly thicker sheets, but the ratio of half-thickness from simulations with different mass ratios remains same. We get very similar behaviour for the current sheets' width and length from the two filterings. So the filtering cut-off does not affect the mass ratio scaling. We further extended our analysis to process data using the Savitzky–Golay filter in place of the FFT filter, with parameters selected to preserve {the key structural features, specifically the peaks and gradients of the current density.} The resulting outcomes are as follows: the average half-thickness of the current sheets detected were found to be $0.31 d_i$, $0.22 d_i$ and $0.16 d_i$ for S0, S1, and S2, respectively, leading to a ratio of 1.4 and 1.37 for S0/S1 and S1/S2, respectively, which is close to $\sqrt{2}$. The obtained values are similar to those observed via FFT-based filtering, underscoring the robustness and consistency of our findings. 

A simulation with higher resolution was also performed for mass ratio $100$, i.e., $S3$. Due to the limitations of computational power, the simulation was run up to $\omega _{p,i}t=967$ cycles. Using the 3D analysis with BFS algorithm, we find an average half-thickness of $0.14d_i$, and corresponding width and length of $0.57d_i$ and $5.20d_i$, respectively. By comparing the results of current sheet dimensions from $S3$ with $S2$ we get very similar results. To check whether our results are affected by particle noise, simulations with higher particles-per-cell(ppc) were also performed, which were run up to $\omega _{p,i}t=1146$ cycles. Table \ref{tab:setups} shows $S4$ and $S5$ simulation runs with $6^3$ ppc per species (a 73\% increase) for mass ratios $50$ and $100$, respectively. $S4$ simulation gives average halfthickness value of $0.22 d_i$ and $S5$ with value of $0.16 d_i$ for BFS. Comparing $S4$, $S5$ with $S1$, $S2$ respectively, very similar results for the dimension of current sheets are found. Similar results were found with DBSCAN. Hence, our results are not significantly affected by particle noise.

\section{Scale dependence}
\subsection{Scale dependent Kurtosis}
\begin{figure}
\centering
\includegraphics[width=1 \linewidth]{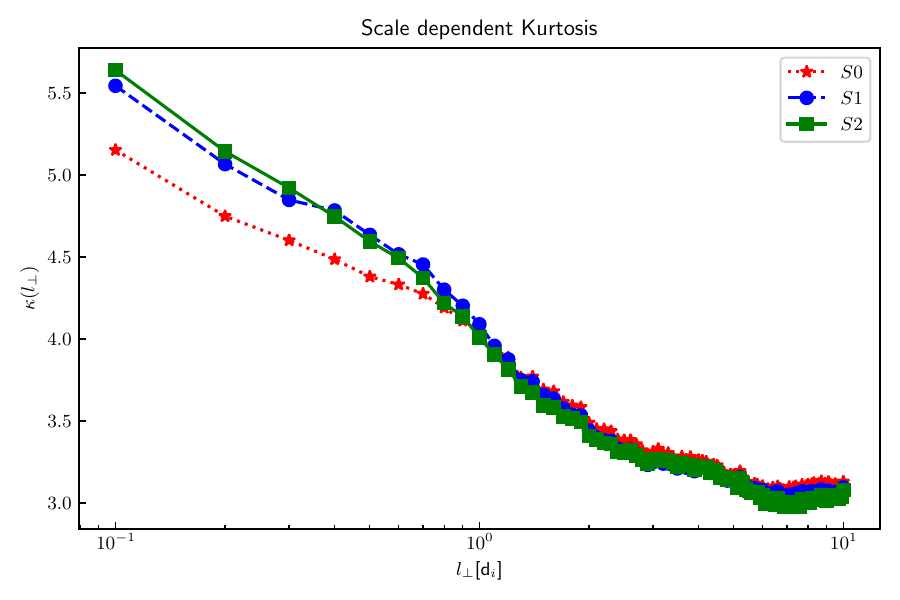}
\caption{ Scale dependent Kurtosis averaged over $\omega _{p,i}t=600-1400$ for $S0$, $S1$ and $S2$.}   
\label{fig:Kurtosis}
\end{figure}

The occurrence of irregular, intense, and spatially localized fluctuations within a turbulent field are denoted by intermittency. These events produce sharp gradients and lead to statistical distributions that deviate from Gaussian behavior. Such behavior is an intrinsic property of turbulent flows, including those in the interplanetary plasma \citep{matthaeus2015intermittency}. In magnetized plasmas, nonlinear interactions naturally generate small-scale intermittent structures—most notably current sheets \citep{Biskamp2003}. Kurtosis is a measure of intermittency in a turbulent flow. Kurtosis tells us about the deviation of the tail of the distribution from a gaussian. Scale-dependent kurtosis is given by 
\begin{center}
    \begin{equation}
        \kappa (l) = \frac{\langle \delta B_l ^4\rangle}{\langle \delta B_l ^2\rangle ^2},
    \end{equation}
\end{center}
where $\delta B_{l}= \mathbf{\hat{{l}}\cdot [B(r+{l})-B(r)]}$ is the increment in magnetic field along the direction $\mathbf{\hat{{l}}}$  and the angled brackets mean spatial averaging \citep{chhiber2018higher}. A Gaussian distribution has a kurtosis of 3, and any value larger than 3 comes from flat tails departing from the Gaussian. In Fig.~\ref{fig:Kurtosis} we plot the kurtosis in perpendicular plane at different length scales for the different simulations. We see that above $l_{\perp}=6d_i$ the kurtosis is around 3, which indicates that the distribution is Gaussian at this scale. At smaller scales, the kurtosis increases to values larger than 3, indicating the presence of localized structures at those scales, consistent with solar wind observations \citep{wu2013intermittent, chhiber2018higher, macek2015themis}. The kurtosis also increases with increasing mass ratio ($\kappa_{S0} > \kappa_{S1} > \kappa_{S2}$) at the smallest lags. This is also consistent with previous studies \citep{edyvean2024scale} where they showed a logarithmic increase in kurtosis at the smallest scales with decreasing electron mass.

\subsection{Scale dependent properties of current sheets}

\begin{figure}
\centering
\includegraphics[width=1\linewidth]{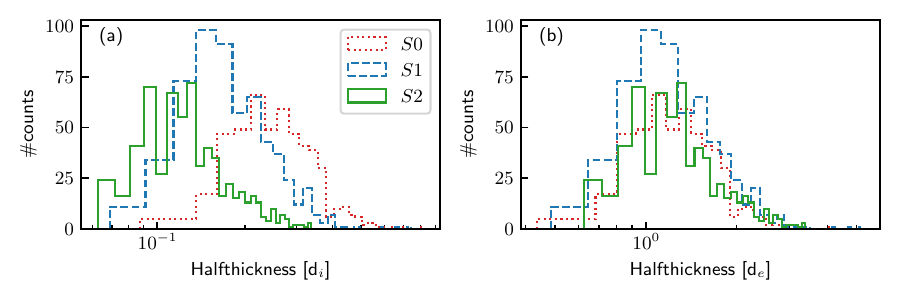}
\caption{ Number of current sheets with different thickness in each bin (a) in units of ion skin depth $d_i$ and (b) in units of electron skin depth $d_e$. Colors red, blue, and green are from simulations $S0$, $S1$, and $S2$ respectively from time step $\omega_{p,i}t=600-1400$.} 
\label{fig:thickness_hist}
\end{figure}

Fig. \ref{fig:thickness_hist} shows the distribution of current sheet half-thicknesses in $S0$, $S1$, and $S2$. The left panel shows the half-thickness in units of $d_i$ and the right panel shows the half-thickness in units of $d_e$.  The distributions for all three simulations overlap perfectly when visualized in units of $d_e$. The peaks of all distributions fall close to 1 $d_e$, indicating that, regardless of the mass ratio, the current sheet thicknesses are comparable to $d_e$.

\begin{figure}
\centering
\includegraphics[width=1\linewidth]{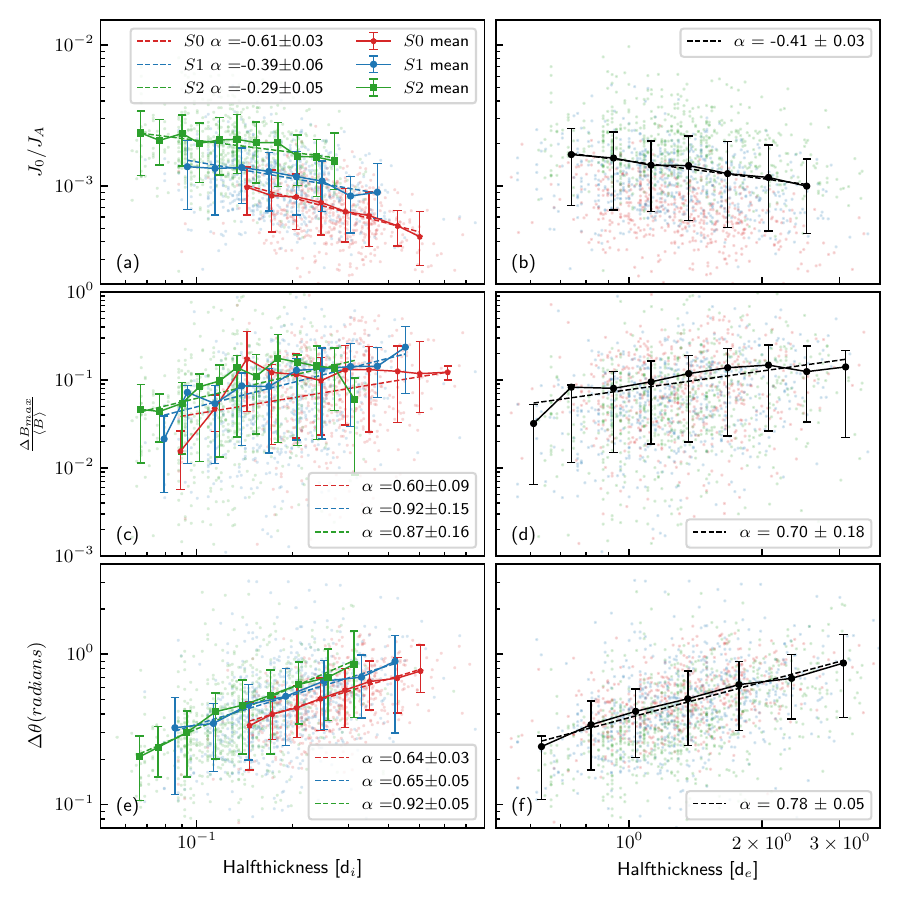}
\caption{ The scale dependence of $J_0/J_A$, $\Delta B_{max}/\langle B \rangle$ and shear angle $\Delta \theta$ is shown in the left panels in units of the ion skin depth $d_i$ and in the right panels in units of the electron skin depth $d_e$. The red, blue, and green points correspond to simulations $S0$, $S1$, and $S2$, respectively, for the time interval $\omega_{p,i}t = 600$--$1400$. The black points represent the mean of the combined data from all three simulations in each bin. The scatter plots display the simulation data, and the mean of the binned data is illustrated with line plots; the dashed lines indicate the linear fits. The error bars on the mean profiles represent the 15th and 85th percentiles of the values within each bin.}
\label{fig:J0_thickness} 
\end{figure}



Fig. \ref{fig:J0_thickness} shows the scale dependent properties of current density $J_0/J_A$(top panel), magnetic field change across the current sheet, $\Delta B_{max}/\langle B \rangle$(middle panel), and shear angle across the current sheet $\Delta \theta$(bottom panel) from different runs, plotted against the half-thickness of the current sheet. These are measured for each identified current sheet. $J_0$ is the average absolute parallel current density in the current sheet and $J_A$ is the Alfv{\'e}n current density given by $J_A=en_0V_A$, where $V_A$ is the average local Alfv{\'e}n speed calculated from the local magnetic field and plasma density ($n_0$) within the current sheet. $\Delta B_{\max}$ is the difference in the magnetic field component that is tangential to the current sheet. To obtain it, we take two points on opposite sides of the sheet across its thickness and extract the magnetic field components that lie tangent to the sheet at these locations, since this is the component that shows the largest variation across the sheet. $\langle B \rangle$ denotes the average magnetic field strength within the sheet. The shear angle, $\Delta \theta$, is defined as the angle between the tangential magnetic field at the two edges of the current sheet, and $d$ represents the sheet thickness. To compute $\Delta \theta$, we again select the two points across the thickness of the sheet, extract the corresponding tangential magnetic field components, and evaluate the angle between these vectors.

Left panels of Fig. \ref{fig:J0_thickness} show the variation with half-thickness of the current sheet in units of ion-skin depth, and the right panels show the variation in units of electron-skin depth. The scatter plot shows the simulation data in both panels. The dashed lines in the left panel are the linear fitted power laws and the black dashed line in the right panel shows the linear fitted power laws for the combined data of three simulations. The current sheets are sorted into logarithmic bins according to their thickness and the mean value for each bin is shown by the solid markers  with the error bars corresponding to the 15th and 85th percentiles within that bin.
The top row of Fig. \ref{fig:J0_thickness} shows that current sheets with larger current density have smaller thickness. The left panel of the Fig. shows similar behaviour for the three simulations with different scalings. Fig.~\ref{fig:J0_thickness}(a) shows that $J_0/J_A$ is negatively correlated with current sheet thickness and shows a scaling behaviour as 
\begin{equation}
J_0/J_A\propto  (d/d_i)^{-0.61 \pm 0.03},
J_0/J_A\propto (d/d_i)^{-0.39 \pm 0.06},
J_0/J_A\propto (d/d_i)^{-0.29 \pm 0.05},
\end{equation}
for $S0$, $S1$ and $S2$ simulations respectively. Fig. \ref{fig:J0_thickness} right panel shows the graph normalized in units of electron skin depth $(d_e)$. We see the three simulations overlap with the scale dependence given by $J_0/J_A\propto  (d/d_e)^{-0.41 \pm 0.03}$.

Fig. \ref{fig:J0_thickness}(middle panel) shows that $\Delta B_{max}/\langle B \rangle$ is positively correlated with current sheet thickness. Fig.~\ref{fig:J0_thickness}(c) shows the different scaling behaviour of mean profile of the binned data with the best fit as
\begin{equation}
\Delta B_{max}/\langle B \rangle \propto (d/d_i)^{0.60 \pm 0.09}, 
\Delta B_{max}/\langle B \rangle \propto (d/d_i)^{0.92 \pm 0.15},
\Delta B_{max}/\langle B \rangle \propto (d/d_i)^{0.87\pm 0.16}
\end{equation}
for $S0$, $S1$ and $S2$ simulations respectively. Also $\Delta B_{max}/\langle B \rangle$ variation with thickness of current sheets normalised to $d_e$ is given by $\Delta B_{max}/\langle B \rangle \propto (d/d_e)^{0.70 \pm 0.18}$ for the mean profile of the binned data, Fig. \ref{fig:J0_thickness}(d).

Fig. \ref{fig:J0_thickness}(bottom panel) shows that there is a positive correlation between the shear angle $\Delta \theta$ and the thickness of the current sheets. When current sheet thickness is normalized to ion-skin depth $(d_i)$, the best fit for the mean profile of the binned data Fig. \ref{fig:J0_thickness}(e) of $S0$, $S1$ and $S2$ simulation respectively is 
\begin{equation}
\Delta \theta \propto (d/d_i)^{0.64\pm 0.03},
\Delta \theta \propto (d/d_i)^{0.65\pm 0.05},
\Delta \theta \propto (d/d_i)^{0.92\pm 0.05}
\end{equation}
In units of electron skin depth the variation is given by Fig. \ref{fig:J0_thickness}(f) with $\Delta \theta \propto (d/d_e)^{0.78\pm 0.05}$ for the binned mean values. The scale dependence of these quantities can be expressed as $A \propto  (d/d_s)^{\alpha}$ where $A$ represents one of the quantities $J_0/J_A$, $\Delta B_{\max}/\langle B \rangle$, or the shear angle $\Delta \theta$. Here, $\alpha$ denotes the corresponding power-law index, and $s$ indicates the particle species, with $s = e$ for electrons and $s = i$ for ions. The values of the power-law indices $\alpha$ for all cases are summarized in Table ~\ref{tab:power_law}.

\begin{table}
  \begin{center}
\def~{\hphantom{0}}
  \begin{tabular}{lcccccccc}
 Name & Length normalization & $J_0/J_A$ & $\Delta B_{max}/\langle B \rangle$ & $\Delta \theta$ \\[10pt]

 S0 & $d/d_i$ & $-0.61\pm 0.03$ & $0.60\pm 0.09$ & $0.64\pm 0.03$ \\ 

 S1 & $d/d_i$ & $-0.39\pm 0.06$ & $0.92\pm 0.15$ & $0.65\pm 0.05$ \\
 
 S2 & $d/d_i$ & $-0.29\pm 0.05$ & $0.87\pm 0.16$ & $0.92\pm 0.05$ \\

 Combined & $d/d_e$ & $-0.41\pm 0.03$ & $0.70\pm 0.18$ & $0.78\pm 0.05$ \\
 
  \end{tabular}
  \caption{Power law indices $\alpha$ for scale-dependent properties of quantity $A$ with the form $A\propto (d/d_s)^{\alpha}$.}
  \label{tab:power_law}
  \end{center}
\end{table}

Similar scale dependencies were found in the solar wind by \citet{vasko2022kinetic} from Wind spacecraft observations at $1$ AU and \citet{lotekar2022kinetic} from Parker Solar Probe observations around $0.2$ AU from the sun. They claimed that since the scale dependence of current density, amplitudes, and shear angles are similar to those expected from the turbulent fluctuations, these kinetic scale current sheets are formed by turbulent fluctuations in the solar wind. Also, from \citet{vasko2022kinetic} eqns. $10$ and $11$, the expected power law from turbulent fluctuations for $J_0/J_A$ is between $-0.1$ and $-2/3$, and for $\Delta B_{max}/\langle B \rangle$ is between $1/3$ and $0.9$ at scales above and below proton kinetic scales. Our results are well within these range of values and give spectral slopes in agreement with \citet{vasko2022kinetic} and \citet{lotekar2022kinetic}. These observations support the assertion that sub-ion scale current sheets are formed in the solar wind by turbulent fluctuations.

A scatter plot between $\Delta B_{max}/B$ and $\Delta\theta$ (not shown here) shows that most points satisfy $\Delta B_{max}/B \le \Delta\theta$, with approximately $73\%$ of the data lying within the range $0.9\,\Delta\theta < \Delta B_{max}/B < 1.1\,\Delta\theta$. This indicates that the magnetic-field change across the current sheets is primarily driven by rotation rather than by a change in the field magnitude. This behaviour is characteristic of rotational discontinuities (RDs), in which the change in magnetic field is mainly due to rotation and its magnitude remains nearly unchanged \citep{tsurutani2018review}. Moreover, the power law slopes of $\Delta B_{max}/\langle B \rangle$ and $\Delta\theta$ are found to be very similar, suggesting that this rotational behavior persists down to electron scales. Similar results have been reported in \citet{vasko2022kinetic} and \citet{lotekar2022kinetic}, where the power-law scalings of both $\Delta\theta$ and $\Delta B_{max}/\langle B \rangle$ were also very similar.

\subsection{Volume and dissipation fractions}
\begin{figure}
\centering
\includegraphics[width=1\linewidth]{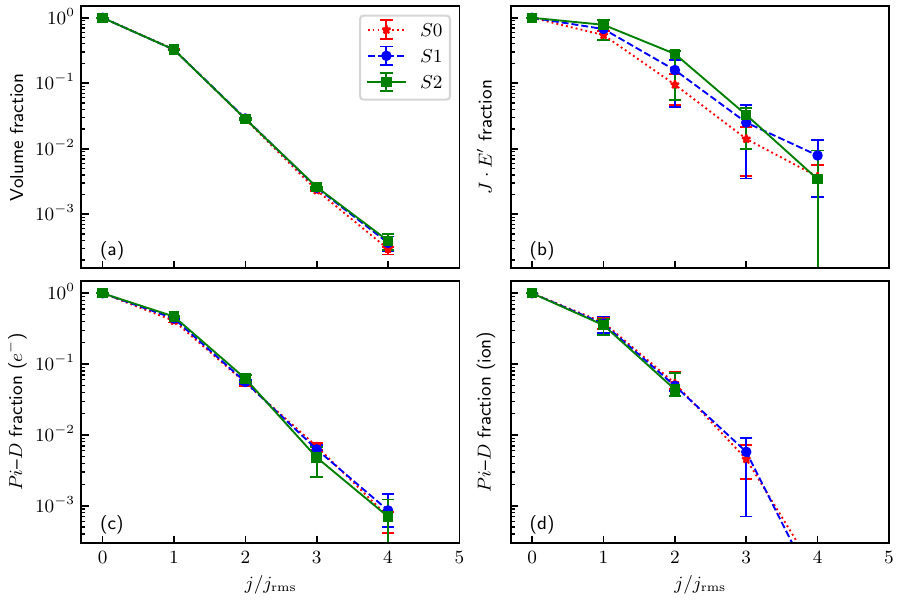} 
\caption{ The fraction of volume occupied by the current density (a), the $\mathbf{J.E'}$ fraction (b), and the $Pi-D$ fraction for electrons (c) and ions (d) are shown, averaged over the interval $\omega_{p,i}t = 600-1400$. Red, blue, and green curves correspond to simulations $S0$, $S1$, and $S2$, respectively. The error bars of the mean profiles represent the 25th and 75th percentiles of the quantities within each bin.}
\label{fig:jdotE_pid_subplots}
\end{figure}

The rate by which the electromagnetic energy is converted into kinetic energy is the work done by the electromagnetic field given by Zenitani dissipation measure~\citep{zenitani2011new}, $\mathbf{J.E'}$, where $\mathbf{J}$ is the current density and $\mathbf{E'}$ is the non-ideal electric field given by $\mathbf{E'=E + v_e\times B}$, \citep{bandyopadhyay2023collisional}.
To understand the dissipation in more detail we obtain the pressure strain interaction for species $a$ given by
\begin{center}
\begin{equation}
    Q^{(a)}=-\mathbf{(P_a.\nabla ).u_a}
\end{equation}
\end{center}
It can be further decomposed into $Pi-D$ and $p\theta$ which are compressible and incompressible contributions respectively (however see \citet{adhikari2025helmholtz}). $Q$ is written as
\begin{center}
\begin{align}
    Q &= -\Pi_{ij} D_{ij} - p\, \nabla \cdot \mathbf{u}, \label{Qpid}
\end{align}
\end{center}
where $\Pi_{ij} = P_{ij} - p\, \delta_{ij}$ and,
\begin{center}
\begin{align}
    \quad D_{ij}=\frac{1}{2}(\partial u_i/\partial x_j + \partial u_j/\partial x_i)- \frac{1}{3}\mathbf{(\nabla \cdot u)}\delta _{ij}. 
\end{align}
\end{center}
$D_{ij}$ is traceless velocity gradient and $\Pi _{ij}$ is traceless deviatoric pressure, $p=P_{ii}/3$ is the scalar pressure. Here Einstein summation convention is used. The first term on right-hand-side of Eq.~\ref{Qpid} is referred to as the $Pi-D$ term, while the second term is denoted as the $p\theta$ term.

We first obtain the fraction of volume occupied by current densities above a particular threshold $j_{thr}$ in Fig. \ref{fig:jdotE_pid_subplots}(a). We see that the volume occupied by current densities is $1.0$ for $j/j_{rms}\ge 0$ and then it decreases to $32\%$ for current densities $j\ge j_{rms}$ for all three simulations.
The probability density functions (PDF) of $\mathbf{J.E'}$ for the three simulations $S0$, $S1$, and $S2$ (not shown here) show a highly non-Gaussian flat tail indicating intermittent dissipation. We calculate the fraction of $\mathbf{J.E'}$ and $Pi-D$ for ions and electrons in our simulations. Fig. \ref{fig:jdotE_pid_subplots}(b) shows the plot of the $\mathbf{J.E'}$ fraction contributed by current densities above a particular $j/j_{rms}$. Similarly, Fig. \ref{fig:jdotE_pid_subplots}(c) and (d) show the $Pi-D$ fraction of electrons and ions, respectively, for the three simulations.
Fig. \ref{fig:jdotE_pid_subplots}(b) shows that for $j/j_{rms}\ge 0$, the $\mathbf{J.E'}$ fraction is $1.0$; as we go to higher current densities, the fraction drops. For $j\ge j_{rms}$, the fraction is $53\% $, $66\%$ and $78\%$ for $S0$, $S1$ and $S2$ simulations respectively. The $\mathbf{J.E'}$ fraction occupied by the simulation with a higher mass ratio is larger than the lower mass ratio up to $j=3j_{rms}$. This means that the higher mass ratio simulation dissipates more strongly in the regions of higher current density. Similar trends are seen for $Pi-D$ fraction for electrons and ions. Fig. \ref{fig:jdotE_pid_subplots} (d) shows around $35\%$ of the $Pi-D$ ion fraction occupied by current sheets with $j\ge j_{rms}$ in $S0$, $S1$ and $S2$ simulations. The fraction further decreases for higher current density at the same rate for the three simulations. The $Pi-D$ fraction for electrons in Fig. \ref{fig:jdotE_pid_subplots} (c) shows similar trend with $Pi-D$ electron fraction of $40\%$, $44\%$ and $47\%$ for $S0$, $S1$ and $S2$ simulation respectively for $j\ge j_{rms}$. We observe that the mass ratio change has little effect on the conditional $Pi-D$ distribution, however there are caveats. $Pi-D$ is most intense near current sheets rather than on top of the current sheets and this can be captured by lagged correlations \citep{parashar2016propinquity}. It has also been shown by \citet{edyvean2024scale} that the average $Pi-D$ and $p\theta$ for the whole simulation changes significantly with mass ratio, with realistic mass ratios having significantly more contributions from $p\theta$ compared to lower mass ratios. We find that for $S2$ simulation the $\mathbf{J.E}'$ fraction for $j \ge 2j_{rms}$ is approximately $28\%$, while the $Pi-D$ fractions for ions and electrons are about $4\%$ and $6\%$, respectively. This indicates that the $\mathbf{J.E}'$ contribution dominates the dissipation in regions of high current density, compared to the $Pi-D$ fraction. Furthermore, for $j\ge j_{rms}$ the ion $Pi-D$ fraction is weaker than the corresponding electron contribution, suggesting that electrons dissipate more strongly than ions in these regions. Similar trends are shown by $S0$ and $S1$ simulations also.


\section{Conclusions}
We have performed 3D kinetic particle in cell simulations, using KAW eigenvector relations for initialization of simulations, in order to understand the turbulent fluctuations at sub-ion and electron scales. KAW modes are proposed as the possible wave modes in the kinetic range turbulence of solar wind. The ion-to-electron mass ratio  is varied keeping the same plasma $\beta$ and also $T_i\approx T_e$. This setup allows us to clearly investigate the role of waves, ion scales, electron scales, intermittency, and dissipation in this turbulence. 



The perturbation ratios of the electric and magnetic fields are compared with the theoretical KAW relations as functions of $k_{\perp}$, revealing the dominance of KAWs at sub-ion scales in all simulations. These results are similar to our previous results in \citet{sharma2024kinetic}, but now we have extended them to the more general  $k_{\perp}$ variation, instead of specific wavenumbers, and to higher mass ratios. Resolution and particles-per-cell have been varied to verify the robustness of the results. A more detailed examination of potential damping of KAW fluctuations for fully kinetic simulations with realistic mass ratio require significantly more computational resources and are left for future studies.

The current sheet structures that form in our simulations are analyzed by two algorithms, namely BFS and DBSCAN. PCA and convex hull have been used to find the length, width and thickness of these sheets. Similar results of the dimensions of current sheets were obtained from both BFS and DBSCAN algorithms. We also tested our results with different amounts of FFT filtering and with Savitzky–Golay filter and we get similar results. The thickness clearly shows an inverse scaling with the square-root of the ion-to-electron mass ratio (especially in the 3D structure analysis), indicating a thickness of the electron skin-depth. However, because of the choice of keeping $\beta$ close to unity in this work, the electron gyro-radius also has the same scaling. Further studies will be needed to tease out the difference between the electron gyro-radius and skin-depth physics. This can give information on the mechanism limiting the thinning of the current sheets. Another limitation of this work is the reduced mass-ratio, which leaves open the question whether the KAWs cascade down till the realistic electron scale. These questions are open for future studies.

In our simulations the number of current sheets remains approximately unchanged up to $\omega _{p,i}t=1800$, which suggests that tearing instability may not be breaking them up in our simulations. However, previous fully kinetic simulations as well as two-fluid 10-moment simulations of collisionless plasmas have shown that thin current sheets can become unstable to the tearing mode as well as the plasmoid instability \citep{daughton2006fully, daughton2011role, liu2013bifurcated, edyvean2024scale}. The widths are calculated using the PCA direction and the lengths are calculated using the convex hull, which is a faster version of simply finding the farthest points. This ignores the curvature of the sheets \citep{serrano2025scale}. Nevertheless, it gives some idea of the scaling of the widths and thickness. {The widths show a decrease with increasing mass-ratio, but it is weaker than the square-root of mass-ratio scaling}. If the ratio of widths to thickness increases with mass-ratio, it hints that the tearing instability may come into picture at realistic mass-ratios \citep{tsareva2024fast}. The lengths show even weaker dependence on the mass-ratio. This indicates that at realistic mass-ratios, a kinetic kink instability could come into play \citep{suzuki2002current}. Further investigation is needed to find out whether such kinetic instabilities play a role in shaping these structures and also whether there is reconnection and electron scale jets coming from these electron thickness current sheets.




Examining the scale-dependent kurtosis across different simulations, we observed higher intermittency at sub-ion and electron scales, with the largest mass ratio simulation exhibiting the highest intermittency. We also studied the scale-dependence of current density $J_0/J_A$, magnetic field variation $\Delta B_{max}/\langle B \rangle$ and shear angle $\Delta \theta$ for different simulations. When these quantities are plotted with current sheet thickness normalised to ion skin depth $(d_i)$ we see that simulations with different mass ratios are separated, but when plotted in electron skin depth $(d_e)$ units, we see that the three simulation results overlap indicating that current sheets are formed at similar electron scales. The scale-dependence of these properties are well within the expected scaling for the turbulent fluctuations and our results are quite similar to the results from Wind spacecraft and Parker Solar Probe observations shown by \citet{vasko2022kinetic} and \citet{lotekar2022kinetic}. We find that the tangential magnetic field change across the current sheets is mostly rotational, and the similar power laws of $\Delta \theta$ and $\Delta B/B$ indicate that this trend should continue down to the electron scales.

The fraction of dissipation by $\mathbf{J.E'}$ and $Pi-D$ for ions and electrons was also obtained for different simulations. In the case of the $Pi-D$ fractions, electrons contribute slightly more than ions in regions of strong current sheets. The $\mathbf{J.E'}$ fraction shows a stronger dissipation for higher mass ratio simulations in the region of higher current density regions. We found that points with $J\ge 2J_{rms}$ roughly occupy 3\% of the domain across the three different mass-ratio simulations. However, in the mass ratio 100 case, these points contribute almost 30\% to $\mathbf{J.E'}$ dissipation, and this fraction is seen increasing with the mass ratio. This shows that at realistic mass ratio, dissipation in these thin electron-scale sheets will be very important. This study reveals important properties of KAW turbulence in the sub-ion range of plasma turbulence. It shows that the electron scale intermittency and dissipation processes will be critical in understanding this regime.

\section*{Acknowledgements}
The authors thank the support and the resources provided by the ``PARAM Seva'' facility at Indian Institute of Technology, Hyderabad.

\section*{Funding}
We would like to thank DST – FIST(SR/FST/PSI-215/2016) for the financial support, along with the Indian Institute of Technology, Hyderabad seed grant and travel grant. We acknowledge support from ISRO-RESPOND grant No. ISRO/RES/2/448/24-25. J.S. thanks the Ministry of Education for the research fellowship. S.S. was supported by Science
\& Engineering Research Board (SERB) grant SRG/2021/001439.

\section*{Declaration of Interest}
The authors report no conflict of interest.


\bibliographystyle{jpp}

\bibliography{jpp_ref}

\end{document}